\newcommand{\switchA}[2] {#2}   

\documentclass[twocolumn,showpacs,prb]{revtex4}
\usepackage{graphicx}

\newcommand{\myfig}[2] {
\begin{figure}
\includegraphics[width=8cm]{#1.eps}
\caption{#2}
\label{#1}
\end{figure} }

\newcommand{\myfigx}[3] {
\begin{figure}
\includegraphics[width=#2]{#1.eps}
\caption{#3}
\label{#1}
\end{figure} }

\newcommand{\myfigs}[3] {
\begin{figure}
\includegraphics[width=#2]{#1.eps}
\caption{#3}
\label{#1}
\end{figure} }

\def\be{ \begin{equation} }
\def\ee{ \end{equation} }

\def\vecs{ \vec{S} }
\def\vecm{ \vec{m} }

\def\eeff{ \eta_{\rm eff} }

\begin{document}

\title{\bf
Frozen Quasi-Long-Range Order in the Random Anisotropy Heisenberg Magnet
}
\author{M.~Itakura}
\affiliation{Center for Promotion of Computational Science and Engineering,
Japan Atomic Energy Research Institute,
Taito-ku, Higashiueno 6-9-3, Tokyo 110-0015, Japan\\}
\date{\today}

\begin{abstract}
Extensive Monte Carlo simulations
are used to investigate
the low-temperature properties of the
random anisotropy Heisenberg model,
which  describes the magnetic behavior of 
amorphous rare-earth-transition metal alloy.
We show that the low-temperature phase in
weak-anisotropy region is characterized by
a {\it frozen-in} power-law spin-spin correlation.
Numerical observation of the power-law exponent $\eta$
indicates non-universal behavior.

\end{abstract}
\maketitle

\switchA{}{
\noindent\fbox{\begin{minipage}{\columnwidth}
{\bf Note:}\\
\small
Owing to the page limitation,
some parts of this manuscript are omitted in the published version.
To obtain the published version, see the comments
at the beginning of the TeX file.
\end{minipage}
}}

Power-law correlation, or quasi-long-range order (QLRO),
usually emerges just at the  critical temperature
of phase transition. The notable exception is
the Kosterlitz-Thouless phase of two-dimensional XY model,
in which power-law correlation 
persists down to zero temperature.\cite{KT}
Recently it has been found that
the low-temperature QLRO phase is quite common in
disordered
three-dimensional systems, \cite{qlro-review}
such as the so-called ``Bragg-glass phase''
of impure superconductors, 
\cite{bragg-nature,bragg-review}
nematic phase of liquid crystal in a porous media,
\cite{lq-rg, lq-mc}
and amorphous rare-earth-transition metal alloy.\cite{ram-review}
These systems are described either by random-field or random-anisotropy
spin models.
Harris, Plischke and Zuckermann first used the following 
random anisotropy model to study 
amorphous metal magnet \cite{HPZ}:
\be
H=-J\sum_{\langle ij\rangle} \vecs_i \cdot \vecs_j - D\sum_i (\vec{n}_i \cdot \vecs_i)^2
\label{h-hpz}
\ee
where $\vecs_i$ is a Heisenberg spin on the lattice site $i$ of
simple cubic lattice, 
$J>0$ is a ferromagnetic coupling constant,
$D>0$ is the strength of uniaxial anisotropy,
the former sum runs over all nearest-neighbor pairs,
and $\vec{n}_i$ is a random unit vector which
describes the random anisotropy at the site $i$.
Similar model, in which $\vecs_i \cdot \vecs_j$ is replaced by
$(\vecs_i \cdot \vecs_j)^2$, is used to study disordered liquid crystals.
While an Imry-Ma-type argument \cite{PPR} of model (\ref{h-hpz})
leads to a conclusion that any nonzero anisotropy $D$
destroys long-range magnetic order in three dimensions,
it is predicted by field theoretical analysis 
\cite{qlro-rg,epsilon,qlro-review}
that the model realizes QLRO
at low-temperature and  weak-anisotropy region.
\switchA{}{
The phase transition between this QLRO phase and paramagnetic phase
is studied by $\epsilon$-expansion techniques, \cite{transition}
and it has been found that
there are no renormalization group 
fixed point in a replica-transformed model of (\ref{h-hpz});
usually this is interpreted as a sign of discontinuous
transition, but no experiments report it.
}
Numerical studies of model (\ref{h-hpz}) 
have been restricted to approximate discretized models
or small sizes,
\cite{mc-ising1,mc-ising2,fisch-o12}
and the QLRO phase was numerically confirmed in Ref. \cite{fisch-o12}.
Similar phase was also found in the simulation of disordered liquid 
crystal. \cite{lq-mc} 
The possibility of spin-glass transition in the
strong-anisotropy region has  been 
investigated by Monte Carlo simulations,
but no conclusive result has been obtained.\cite{mc-ising1,mc-ising2}
Experimental studies have shown that
there is a ``spin freezing transition''
below which zero-field cooled and field cooled magnetization
differ.\cite{sg-ex1,sg-ex2} This phase is referred to 
as ``correlated spin glass'' phase,\cite{csg}
owing to the developed spin correlation length 
as opposed to the usual spin glasses.
Similar glasslike behavior was found
in the recent experiment of disordered liquid crystal.\cite{lq-exp}
However, the predicted QLRO has not yet
been reported experimentally.\cite{SANS}

In the present work, we numerically 
investigate the low-temperature
properties of model (\ref{h-hpz}) on an $L\times L \times L$ lattice
for several values of $D/J$ (including infinity)
using the exchange Monte Carlo method.\cite{repex}
Details of the simulation are
summarized in Table \ref{tab-mc}:
We simultaneously simulate $N_r$ identical systems,
assigning different temperature to each one.
At each Monte Carlo step, exchange of states between adjacent temperatures
$T_{i}$ and $T_{i+1}$ is tried and accepted with a
probability $\mbox{min}[1,
\exp\left((E_i-E_{i+1})(1/T_i -1/T_{i+1})\right)]$,
where $E_i$ denotes the energy of the state at
$T_i$. Thus each state random-walks the temperature space. To maximize
the diffusion constant of this process, a set of temperature $\{T_i\}$ is
adjusted so that
$(\langle E_{i+1}\rangle -\langle E_i\rangle)(1/T_{i+1} -1/T_i) \sim -1$.
The average energy can be easily estimated beforehand from simulations of
smaller systems.
Equilibrium is checked by dividing the
measured data into groups and calculating averages for each group,
then discarding the data of initial groups which are not equilibrated.
\switchA{}{
In the Metropolis update of large $D/J$ cases, 
the new spin is chosen with a probability proportional to 
$\exp( D/T (\vec{n_i}\cdot \vecs_i)^2)$ and 
acceptance probability is calculated using only the 
first term in \ref{h-hpz}.  
For small $D/J$ cases, over-relaxation type update is also employed 
.\cite{OverRelax}
}
When measuring the spin-glass order parameter,
three replicas are placed at each temperature and  overlaps
between them are observed.
\switchA{}{
\noindent\fbox{\begin{minipage}{\columnwidth}
{\bf Note:}\\
In the previous versions of this
manuscript, the following observation scheme was
used in some of exchange MC simulation.
This scheme is incorrect, since the marked replica had always been at higher
temperature just before it is observed, and average over such samples
is biased to high-temperature side.\\
{\it 
A special observation scheme is used when we investigate 
a large system at temperature far below the critical point:
a replica is ``marked'' when it visits the highest temperature,
and if a marked replica reaches the lowest temperature,
we store its configuration into memory and ``unmark'' the replica.
After MC steps proportional to $N_R^2$, a large number of 
independent, equilibrium configurations at low temperature are obtained.
}
\end{minipage}}

}
Average over random anisotropy configurations is performed
over $48 \leq N_s \leq 900$ configurations.
Statistical errors are estimated from 
sample-to-sample fluctuations by the Jackknife procedure.
All the simulations were performed on Fujitsu VPP5000
vector processors at JAERI.

\begin{table}[h]
{
\begin{tabular}{|l|l|l|l|l|l|l|l|l|}
\hline
Fig. & $D/J$ & $L$ & $T_{\rm min}/J$ & $T_{\rm max}/J$ & 
$N_R$ & MCS$_I$ & MCS$_O$ & $N_S$ \\ 
\hline
 1&4.0 & 6 & 0.5 & 2.5 & 17 & 12000 & 12000 & 200 \\
\hline
&    & 12& 0.5 & 2.5 & 65 & 22500 & 22500 & 142 \\
\hline
&    & 24& 1.390 & 1.601 & 22 & 6000 & 4000 & 48 \\
\hline
\hline
2 & 4.0 & 24& 0.832 & 1.601 & 91 & 6000 & 4000 & 96 \\
\hline
\switchA{}{
3 & 10.0& 6 & 0.5 & 2.5 & 17 & 12000 & 12000 & 400\\
\hline
&    & 12& 0.5 & 2.5 & 65 & 22500 & 22500 & 100 \\
\hline
&    & 24& 0.969 & 1.930 & 61 & 60000 & 30000 & 80 \\
}
\hline
\hline
3 & $\infty$&8& 0.5 & 2.5 & 17 & 12000 & 12000 & 900 \\
\hline
&      &12&0.5 & 2.5 & 33 & 24000 & 24000 & 600 \\
\hline
&      &20&0.754&1.665&42 & 48000 & 48000 & 176 \\
\hline
\hline
5 & *& 8 & 0.5 & 1.53 to 1.61 &26 & 5000 & 10000 & 320 \\
\hline
      & * & 12& 0.5 &1.53 to 1.61 & 51 & 5000 & 40000 & 200 $-$ 300 \\
\hline
      & * & 16& 0.5 & 1.53 to 1.61 & 76 & 10000 & 70000 & 140 $-$ 200 \\
\hline
\end{tabular}
}
\caption{
Details of the exchange Monte Carlo simulations.
$N_R$, MCS$_I$, MCS$_O$, and $N_S$ 
denote number of replicas placed between
$T_{min}$ and $T_{max}$, 
number of Monte Carlo steps used for thermalization and discarded, 
number of Monte Carlo steps used for observations, 
and number of samples used for averaging over randomness, respectively.
The sign ``*'' refers to all the values of $D/J$ used in Fig. 5.
}
\label{tab-mc}
\end{table}

We observe both magnetization $\vec{m}= L^{-3} \sum_i \vec{S}_i$
and overlap parameter $q= L^{-3} \sum_i q_i$,
where $q_i=\vec{S}^\alpha_i \cdot \vec{S}^\beta_i$ is an overlap at site $i$
between two independent replicas $\alpha$ and $\beta$ which
we run in parallel.
Figure \ref{gb4} shows plots of Binder's cumulant of $\vec{m}$ and $q$:
\be
B_m=\frac{5}{2}-\frac{3[\langle |\vec{m}|^4\rangle]}{2[\langle\vec{m}^2\rangle]^2},
\,\,\,\,\,
B_q=\frac{3}{2}-\frac{[\langle q^4\rangle]}{2[\langle q^2\rangle]^2}
\ee
for the $D/J=4.0$ case, where $\langle \cdots \rangle$ and $[\cdots ]$
denote thermal and sample averages, respectively.
A clear crossing at 
$T \approx 1.423$ can be 
observed for both $B_m$ and $B_q$
which clearly  indicates that there is some kind of order 
at low temperatures
(Note that the transition temperature is only 2\% smaller than
that of $D=0$ case, as was found in Ref.~ \cite{fisch-o12}).
\myfig{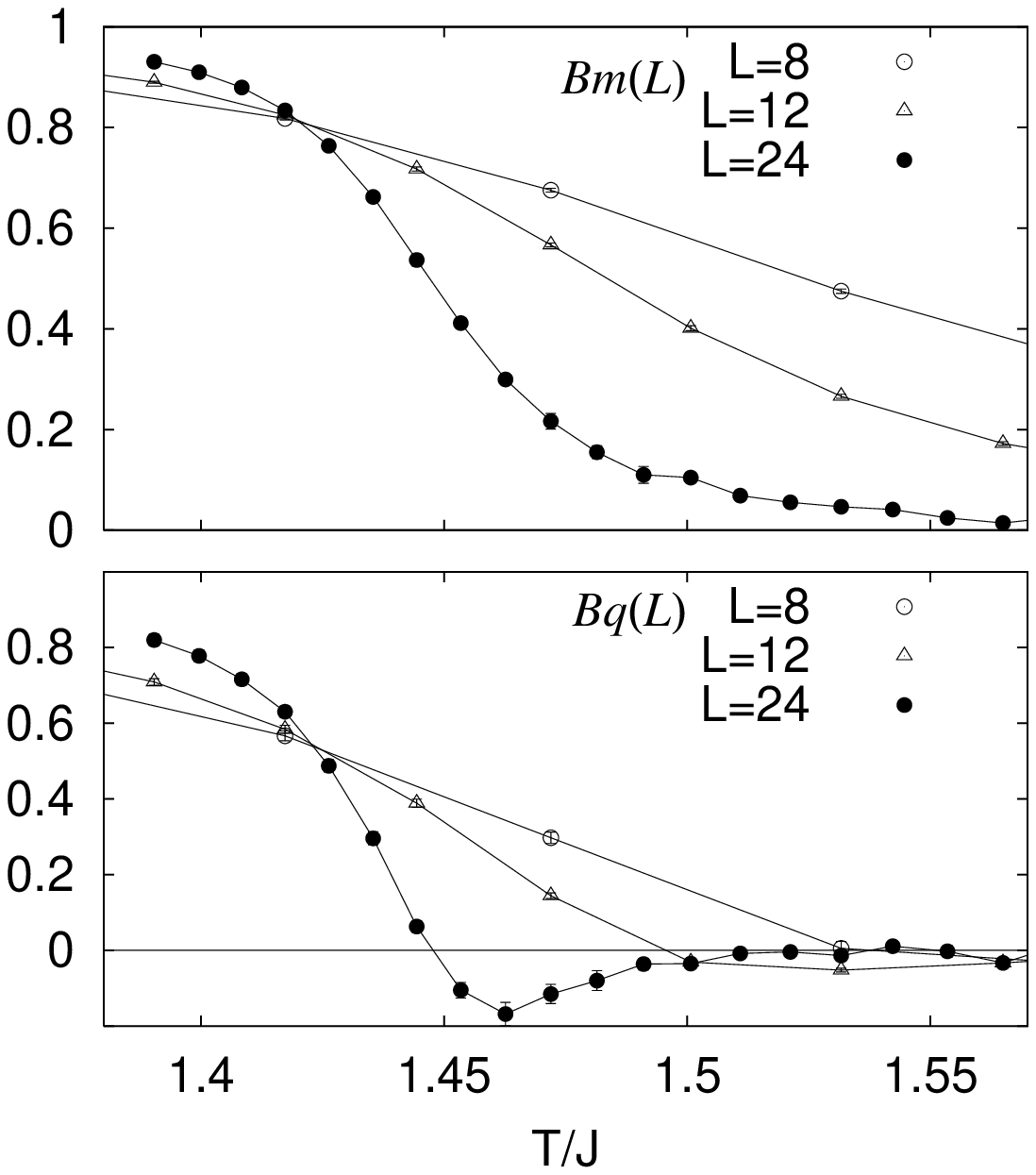}{
Plots of $B_m$ and $B_q$ for the $D/J=4.0$ case.}
To investigate the nature of the ordered phase, we observe the
effective exponent of $\eta$ 
defined between
two different system sizes $L_1$ and $L_2$:
\be
2-d-\eeff =
\frac{\ln [\langle \vecm (L_2)^2\rangle] - \ln [\langle \vecm (L_1)^2\rangle]}
{\ln(L_2)-\ln(L_1)},
\ee
where $d$ denotes spatial dimension (in the present case, $d=3$).
When spin-spin correlation at a distance $r$ is proportional to 
$r^{2-d-\eta}$, then $\eeff$ asymptotically approaches to $\eta$ as the size is
increased, while when the correlation is short ranged, $[\langle \vecm ^2\rangle ]$ is
proportional to $L^{-d}$, therefore $\eeff = 2$ is approached.
Figure \ref{eta-pq} shows plots of $\eeff$ for the $D/J=4.0$ case:
The value of $\eeff$ is almost independent of temperature below the critical point.
This behavior is quite different from that of 
Kosterlitz-Thouless (KT) phase,\cite{KT}
in which $\eta_{eff}$ jumps from 2 to 1/4 at the critical point,
then gradually decreases down to 0 as the temperature is lowered down to $T=0$.
The difference from the KT phase becomes remarkably clear when one sees a 
probability distribution of the overlap parameter
$P(q^\prime)=[\langle \delta(q-q^\prime)\rangle]$.
The inset of Fig. \ref{eta-pq} shows $P(q)$ of two sizes $L=8$ and $12$,
for the
$D/J=4.0$ case, at a very low temperature $T/J=0.6815$
which is  well below the critical point;
the sharp peaks, which become sharper as the size is increased,
indicate that only one kind of state
is dominant in the ordered phase
and there are no critical thermal fluctuations.
From these results,  
one can depict the nature of the low-temperature phase as follows:
the system is
trapped around one of the two ground states $\pm \{\vecs_i^{(0)}\}$,
in which spin-spin correlation decays as 
$\vecs_i^{(0)}\cdot \vecs_j^{(0)} \sim |i-j|^{-\eta-1}$.
At the transition temperature, the system ``freezes'' into this
QLRO state, then the magnetic QLRO and the 
spin-glass order simultaneously emerge.
Baker and Kadanoff suggested similar QLRO phase
for models with finite zero-point entropy, such
as antiferromagnetic Potts models, in which
thermal fluctuation is present even at $T=0$.
They speculated that in these models
the point $T=0$, along with the whole low-temperature
phase, is renormalized to some nontrivial fixed point.
But in the present case, the behavior of $P(q)$ suggests that 
there is only one kind of low-temperature state and 
thermal fluctuation is irrelevant in the low-temperature phase.

\myfig{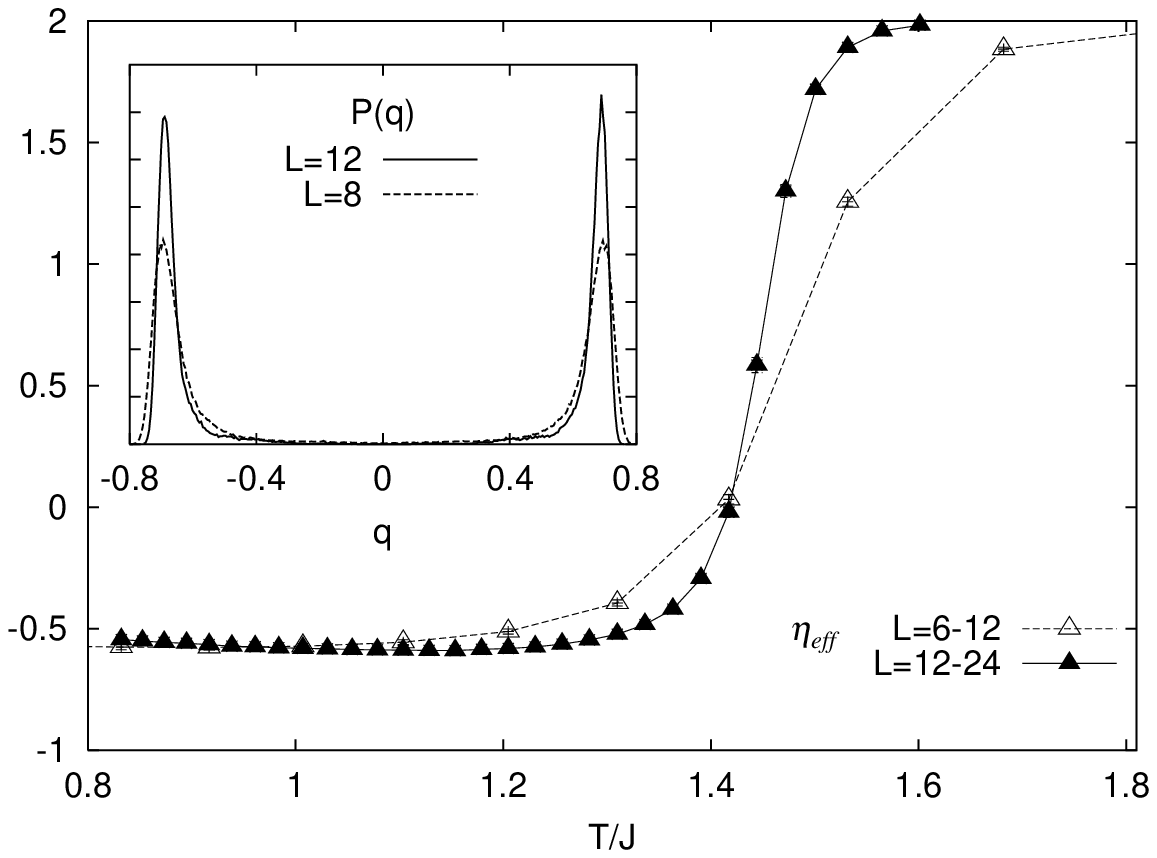}
{Plots of the effective exponent $\eta_{eff}$ against temperature for
the $D/J=4.0$ case. 
Statistical errors are smaller than the size of the symbols.
The inset shows probability distribution of
the overlap parameter $q$ at $D/J=4.0$, $T/J=0.6815$.}

Next, we consider the low-temperature phase in the 
strong-anisotropy region, 
in which QLRO is expected to be destroyed.\cite{mc-ising1,mc-ising2}
Figure 
\switchA{} {\ref{gb10} and }
\ref{gbi} shows plots of $B_m$ and $B_q$
for 
\switchA{the $D/J=\infty$ case, which} 
{ $D/J=10$ and $D/J=\infty$ cases, respectively.The case $D/J=\infty$}
correspond to a kind of Ising spin-glass model
$\sum_{\langle ij\rangle}J_{ij} \sigma_i \sigma_j$ with $J_{ij}=\vec{n}_i \cdot \vec{n}_j$.
The plots of $B_m$ shown in Fig. 
\switchA{}{ \ref{gb10} and }
\ref{gbi}
do not show crossing behavior and are consistent
with the past results
\switchA{.}
{, and indicate that the spin-spin correlation
is already short ranged at $D/J=10$.}
As for the spin-glass order, however,
the crossing of the plots of $B_q$ is rather ambiguous;
they seem to collapse into a line in the low-temperature region.
This result is suggestive of Kosterlitz-Thouless-like
{\it quasi-long-ranged spin-glass order},
but it should be remembered that 
numerical simulations of the three-dimensional
$\pm J$ spin glasses could not exclude
a (false) possibility of KT-like transition, unless recently large-scale
simulations have become possible.\cite{sg-review}
So, in the present work we reserve conclusion on this issue.
The phase diagram shown in Fig. \ref{phd} summarizes these results.
In the weak-anisotropy limit,
magnetic order is either
slowly decaying QLRO or long ranged,
which is difficult to distinguish numerically.

\switchA{}{
\myfig{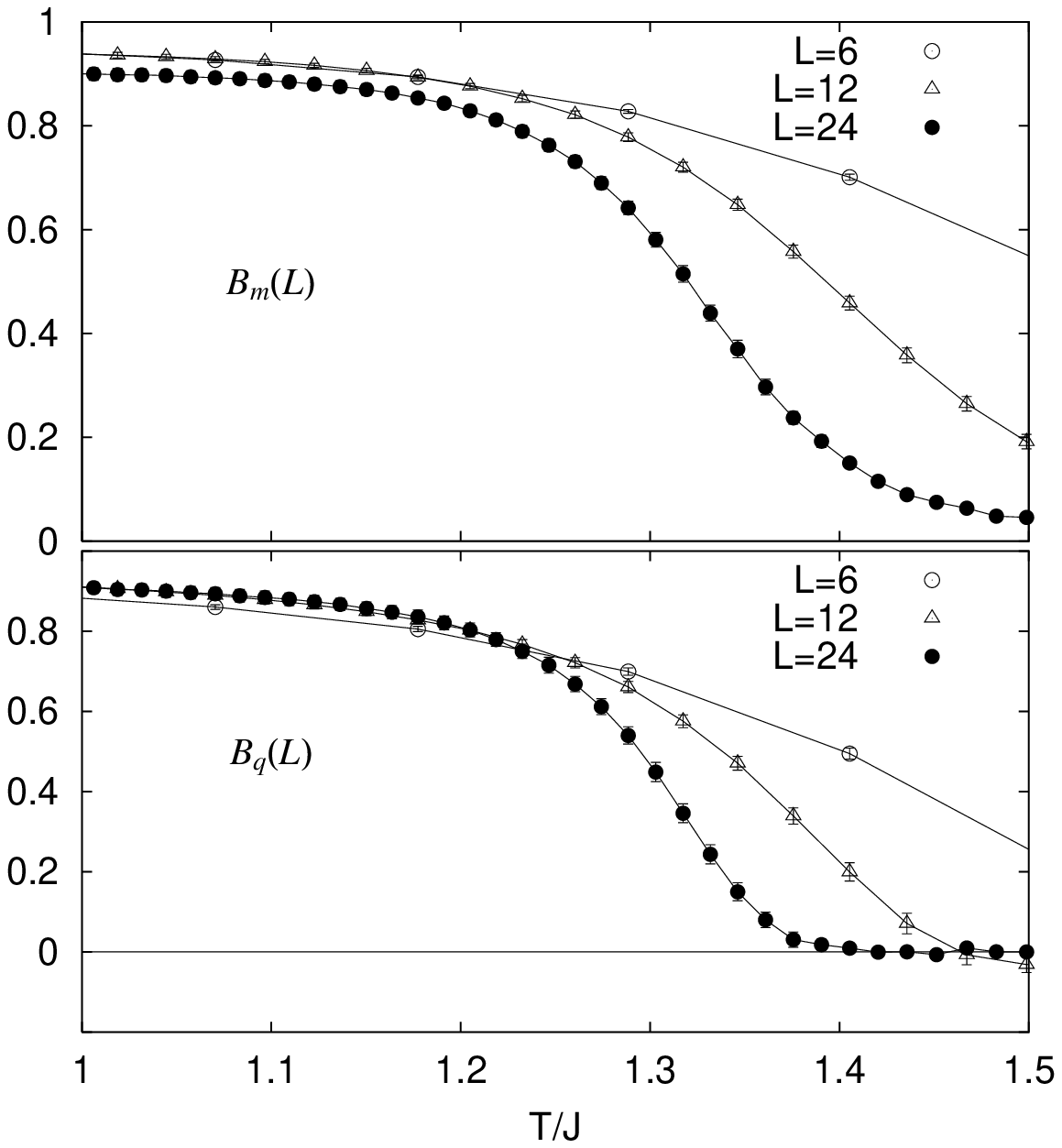}{
Plots of $B_m$ and $B_q$ for the $D/J=10.0$ case.}
}

\myfig{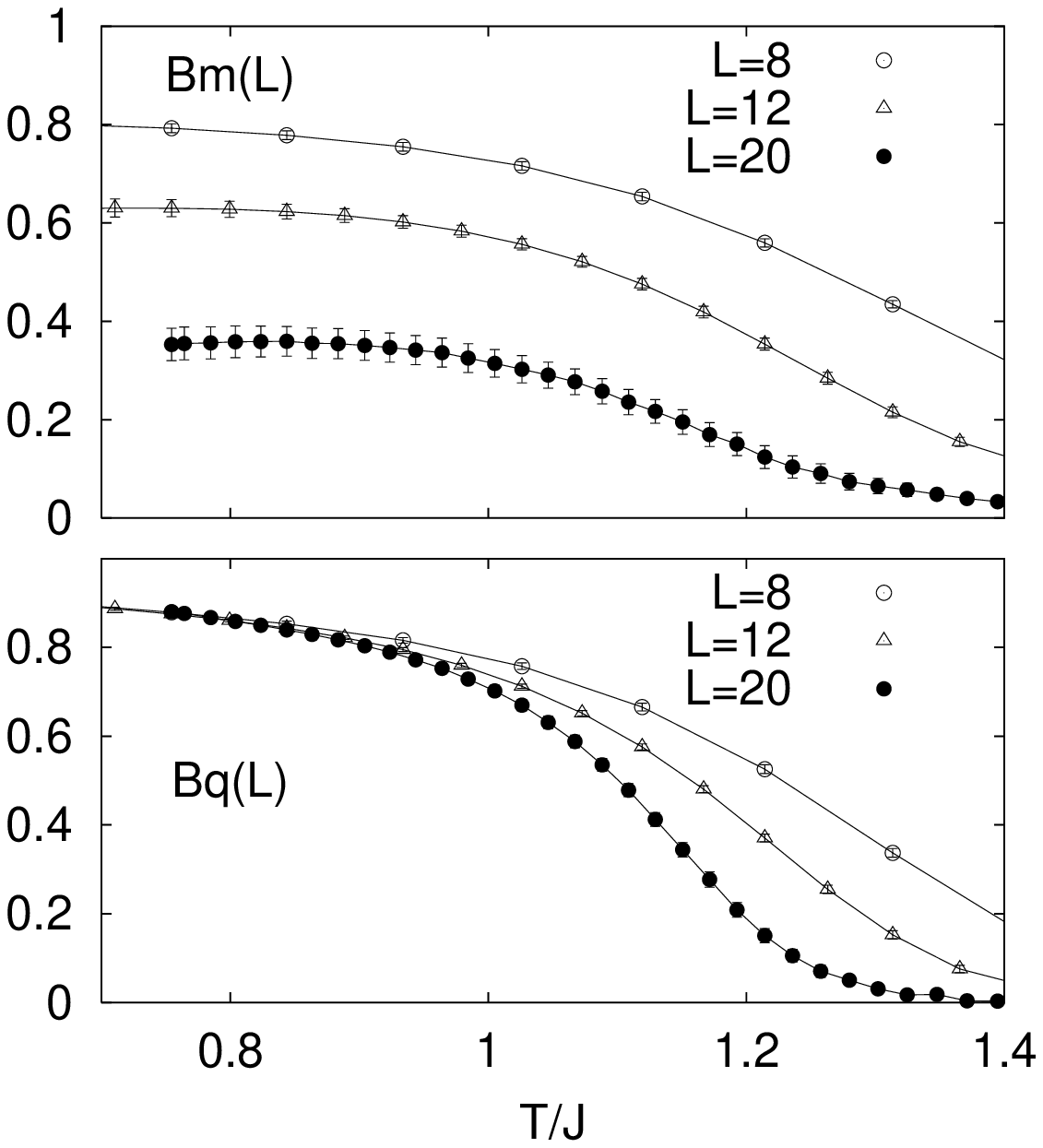}{
Plots of $B_m$ and $B_q$ for the $D/J=\infty$ case.}

\myfigs{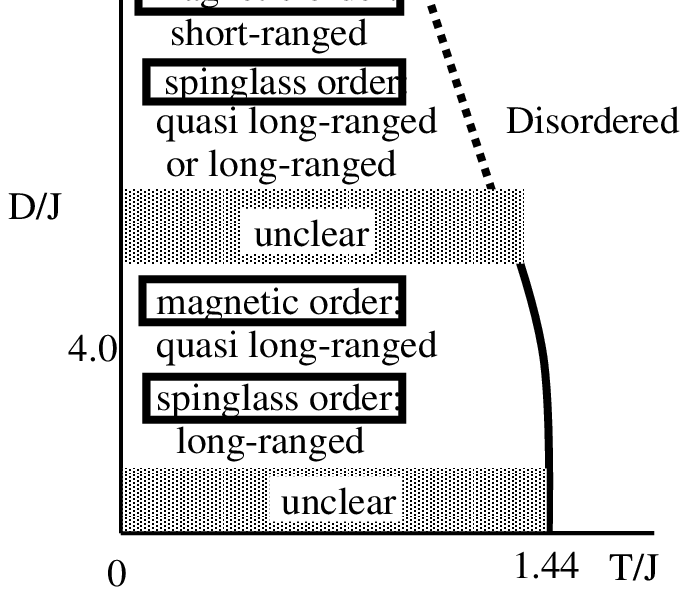}{5cm}{Phase diagram of the random anisotropy magnet,
suggested by the results of the present work.
Dotted lines and shaded area are ambiguous phase boundaries which are hard to
identify by numerical simulations.}

\begin{figure}
\includegraphics[width=3.5cm]{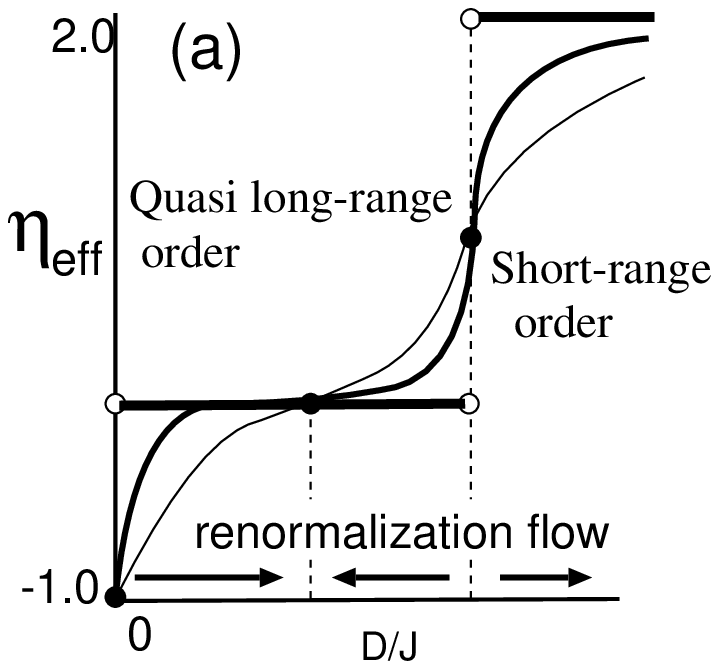}
\includegraphics[width=4.8cm]{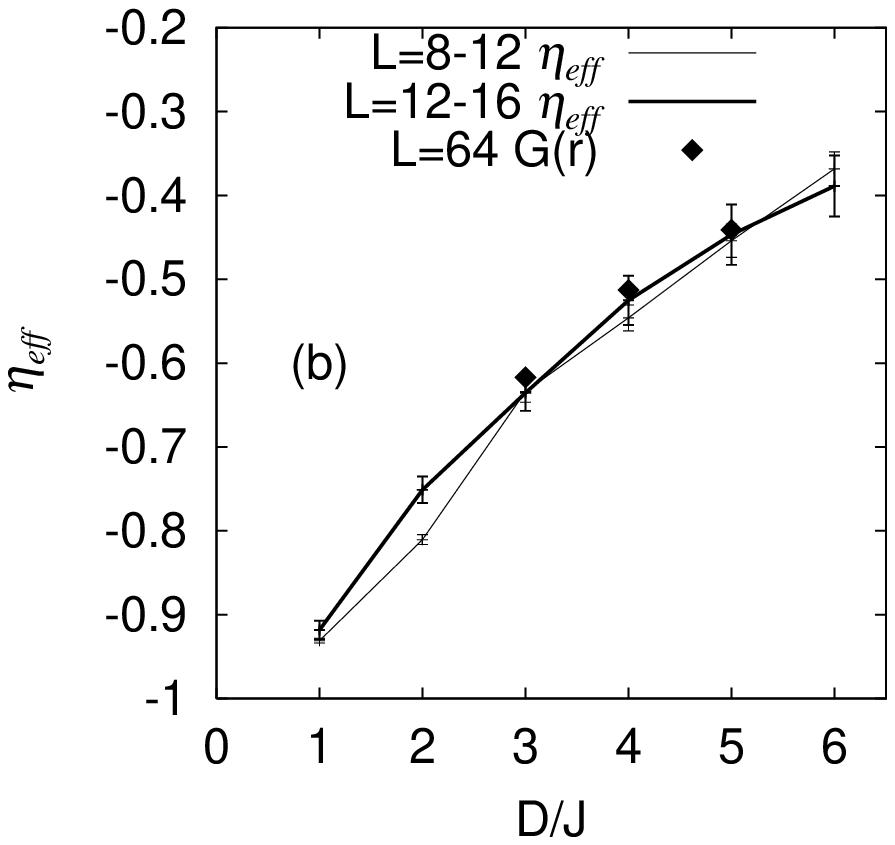}
\caption{
(a) Schematic representation of
finite-size behavior of the effective exponent
$\eeff$ expected from renormalization-group analysis.
Thin and thick curves are finite-size values of smaller and larger
systems, respectively. The bold straight lines show the infinite-volume
limit. 
(b)
Effective exponent $\eeff$ of several values of $D/J$, measured at
$T/J= 0.5$ for all cases. The data represented by a dotted line
are calculated from correlation function of spin configurations
obtained by simulated annealing of $L=64$ system.}
\label{rgmc}
\end{figure}

Now let us consider the following question: Is the exponent $\eta$
universal all over the QLRO phase?
The field theoretical $4-\epsilon$ expansion analysis
\cite{qlro-review,epsilon}
suggests that a stable fixed point at $D/J=O(\epsilon)$ governs
the nature of the whole QLRO phase.
Figure \ref{rgmc}(a)
schematically depicts the finite-size
behavior of $\eta_{eff}$ expected from this theoretical
framework. 
Although the $4-\epsilon$ expansion analysis only foretells 
the presence of a stable
RG fixed point, there should be another, an unstable fixed point
since the QLRO is destroyed in the strong-anisotropy limit.
When the anisotropy ratio $D/J$ 
is smaller than a threshold (unstable fixed point),
it is attracted to a
stable fixed point as it is renormalized, while it will diverge when
larger than the threshold. 
\switchA{}{
Thus plots of $\eta_{eff}$ against $D/J$  at a low temperature
for different sizes will intersect at two fixed points.
}
Figure \ref{rgmc}(b)
shows plots of $\eta_{eff}$ measured at a very low
temperature $T/J=0.5$ (far below the transition line $T\sim 1.4$)
for several values of $D/J$ and for different sizes. 
In a region $3.0 \leq D/J \leq 6.0$,
the two plots
coincide within statistical errors, unlike Fig. \ref{rgmc}(a).
This suggests a nonuniversal QLRO phase in which
the asymptotic value of $\eta$ continuously changes with 
varying $D/J$, or at least that
the renormalization-group flow is very slow.



To see if the observed nonuniversal QLRO  is asymptotic or not,
we investigate larger system $L=64$ with simulated annealing (SA)
method.\cite{SA} The apparent drawback of SA
is that it does not guarantee equilibration, therefore
obtained configuration may not be the true ground state.
Energy of the obtained states is given in Table \ref{tab-e0}.
\begin{table}[h]
\begin{tabular}{|l|l|l|l|l|l|l|}
\hline
$D/J$ & 3.0 & 4.0& 5.0 & 6.0 & 8.0 & 10.0 \\
\hline
$E_0$ &
$-4.7099$ &
$-5.5617$ &
$-6.4705$ &
$-7.4082$ &
$-9.3314$ &
$-11.2871$ \\
\hline
$\Delta E_0$
& 
$\pm 0.0004$ &
$\pm 0.0004$ &
$\pm 0.0005$ &
$\pm 0.0005$ &
$\pm 0.0007$ &
$\pm 0.0005$ \\
\hline
\end{tabular}
\caption{
Energy per spin of the annealed states,
averaged over randomness.
The third row shows standard deviation
of sample-to-sample
fluctuations.
}
\label{tab-e0}
\end{table}

Using the configurations obtained by SA,
we calculate spin correlation function
defined as 
$G(r) \equiv  (3 L^3)^{-1}\sum_{|i-j|=r}
 \vecs_i \cdot \vecs_j $,
where the summation runs over all pairs whose relative position
is either $(r,0,0)$, $(0,r,0)$, or $(0,0,r)$ lattice spacings.
For each random anisotropy configuration, only one annealed
configuration is generated 
and $G(r)$ is averaged over ten random anisotropy configurations.
The summation over lattice sites significantly reduces
statistical errors, and this small number
of random anisotropy configurations
is enough to obtain reasonable precision.
Figure \ref{gr-all} shows plots of $G(r)$ for various values of $D/J$.
For the cases of $D/J \leq 4.0$, $G(r)$ in a range
$4\leq r \leq 16$ is well fitted by a power-law form $G(r)=a r^{-\eta-1}$,
while for $D/J \geq 5.0$ a form
$G(r)=a r^{-\eta-1} \exp(-x/\xi)$ is well fitted.
Owing to the periodic boundary condition, the data in a region $r>16$
deviate from the fitting forms. The values of $\eta$ for
each $D/J$ obtained by the
fitting are shown in Fig. 
\switchA{\ref{rgmc}(b)}{\ref{mc-eta}}, 
and they coincide with the previous
results within statistical errors.
The inset of Fig. \ref{gr-all} shows plot of $\eta$ and $1/\xi$ against $D/J$:
$G(r)$ becomes short ranged at around $D/J  = 5.0$,
therefore the QLRO phase should persist at least up to this 
anisotropy strength.

\myfigx{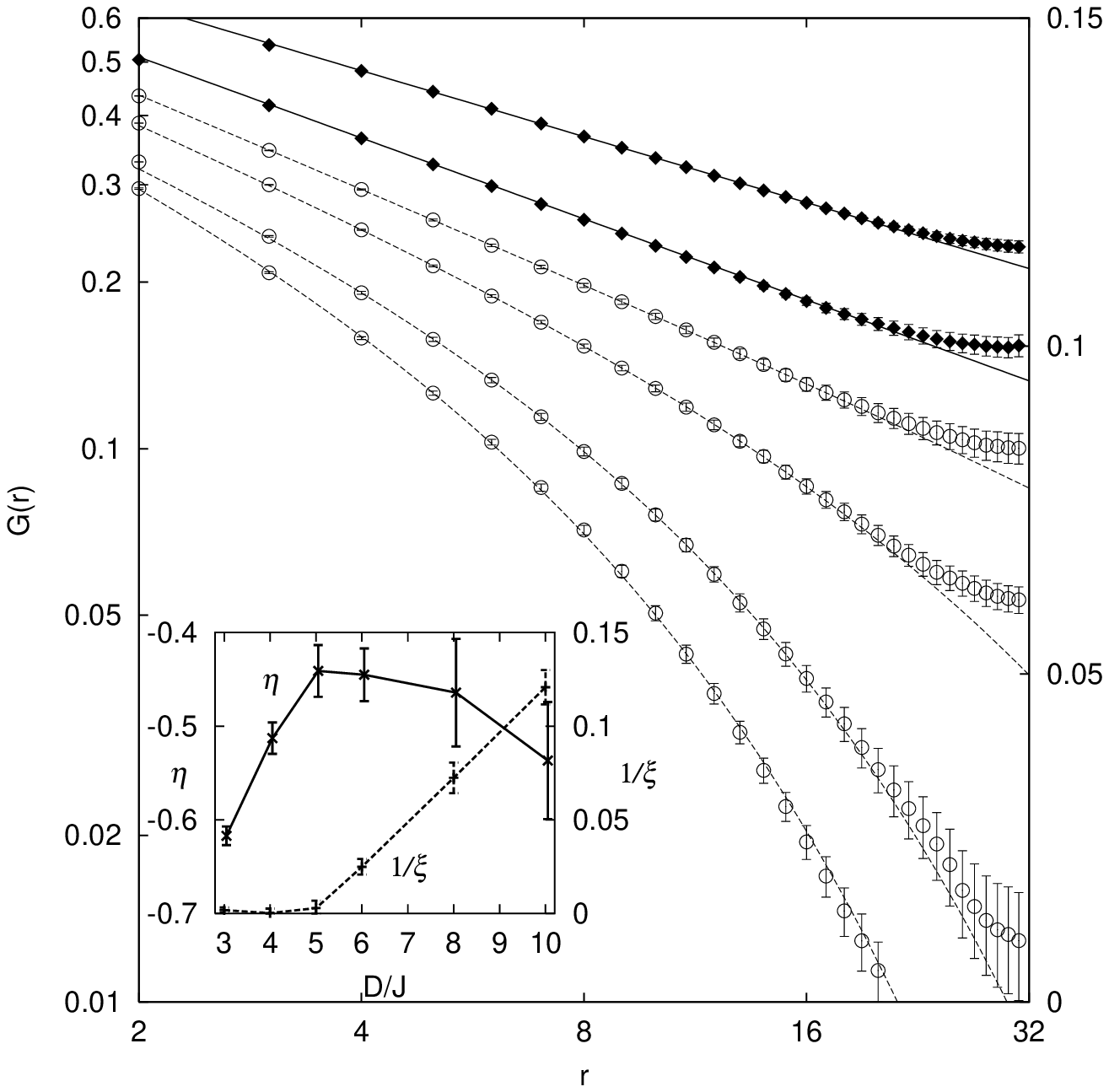}{\columnwidth}{Spin-spin correlation $G(r)$ in the annealed states
for $D/J=3.0$, $4.0$, $5.0$, $6.0$, $8.0$, and $10.0$ cases
(from top to bottom), averaged over ten samples for each case.
Lines are results of fittings, either
$G(r)=a r^{-\eta -1}$ (solid lines) or
$G(r)=a r^{-\eta -1} \exp(-r/\xi )$ (dotted lines).
The inset shows plot of the fitting parameter $\eta$ and $1/\xi$ against $D/J$.
}

Now that we have found a QLRO phase at around $D/J =4.0$,
which is much larger than
that of experimentally realized materials $D/J \sim 1.0$,
a question arises: Why has this QLRO phase never been
observed in the past experiments, such as small angle neutron
scattering \cite{SANS}?
One possible reason may be 
the temperature dependence of the
anisotropy $D$: Generally $D$ increases as the temperature is lowered,
while $J$ remains almost constant.\cite{csg}
Since the ground state is completely modified when $D/J$ changes,
reaching the true ground state requires 
complete reorganization of spins, which is unlikely to happen
no matter how slowly the system is cooled.

In summary, we have carried out extensive Monte Carlo
simulations of the three-dimensional random anisotropy
Heisenberg model and found a QLRO phase 
which is characterized by frozen power-law spin correlations
in the low-temperature, weak-anisotropy region.
This phase persists at least up to the anisotropy strength $D/J \approx 5.0$. 
Finite-size behavior of the effective power-law exponent
in this phase
indicates nonuniversal behavior,
contrary to the field theoretical prediction
of universal exponent.


\end{document}